\documentclass[preprint,showpacs,amsmath,amssymb]{revtex4}
\usepackage{graphicx}

\usepackage{bm}
\usepackage{xy}
\usepackage{epsfig}

\begin{document}
\def\1{\'{\i}}

\title{Factorization procedure and new generalized Hermite functions}

\author{Marco A. Reyes\footnote{Corresponding author.  Electronic mail: marco@fisica.ugto.mx} and M. Ranfer\'i Guti\'errez}
\affiliation{Departamento de F\1sica, DCI Campus Le\'on, Universidad de
Guanajuato \\ Apdo. Postal E143, 37150 Le\'on,Gto., M\'exico}

\begin{abstract}
We propose an alternative factorization for the simple harmonic oscillator hamiltonian which includes Mielnik's isospectral factorization as a particular case.  This factorization is realized in two non-mutually adjoint operators whose inverse product, in the simplest case, 
lead to a new Sturm-Liouville eigenvalue equation which includes Schr\"odinger's equation for the oscillator and Hermite's equation as particular cases for limiting values of the factorization's parameter, and whose eigenfunctions allow us to define new generalized Hermite functions.
\end{abstract}


\maketitle

\section{INTRODUCTION}

One of the most prominent areas of research in mathematical physics in the last 30 years has been that of supersymmetry (SUSY), based on the work of Witten in particle physics.\cite{witten}  In non-relativistic quantum mechanics (QM), the work of Mielnik\cite{bogdan} has been the basis for the development of isospectral potentials for different quantum problems,\cite{sukhatme} and also for higher supersymmetric developments.\cite{sususy}

The best example to introduce SUSY QM is that of the simple harmonic oscillator (SHO), whose hamiltonian
\begin{equation}
H= -\frac{1}{2}\frac{d^2}{dx^2}+\frac{1}{2}x^2
\label{sho}
\end{equation}
possess the eigenfunctions and eigenvalues given by
\begin{equation}
\psi_n(x)=c_n \mbox{H}_n(x) \, e^{-x^2/2} \ , \ \ \ E_n=n+\frac{1}{2} \ ,
\label{eigen}
\end{equation}
where H$_n(x)$ are the Hermite polynomials, satisfying Hermite's equation
\begin{equation}
\mbox{H}_n''(x)-2 x \, \mbox{H}_n'(x)+2n\, \mbox{H}_n(x)=0 \ .
\label{hermit}
\end{equation}
Hermite's equation can be derived from the SHO Schr\"odinger's equation by acting the hamiltonian (\ref{sho}) on the functions (\ref{eigen}) and leaving the equation for the functions H$_n(x)$ alone.
One can find in the literature generalizations of eq.(\ref{hermit}), which define  generalized  Hermite polynomials H$_m(u;t)$,\cite{hermgen2}  H$_n(x;\gamma)$,\cite{hermgen} and H$_n^N(x)$,\cite{relherm} or simply  define the Hermite functions as $h_n(x)$=H$_n(x)\, e^{-x^2/2}$, the quantum eigenfunctions.

The hamiltonian (\ref{sho}) can be factorized using the anhilitation and creation operators
\begin{equation}
a=\frac{1}{\sqrt 2}\left( \frac{d~}{dx}+x \right) , \ \ \ a^*=\frac{1}{\sqrt 2}\left( -\frac{d~}{dx}+x \right) ,
\label{a*a}
\end{equation}
such that $a\, a^*=H+\frac{1}{2}$, and $a^*\, a=H-\frac{1}{2}$.

Mielnik proposed a different way to factorize the hamiltonian, by introducing new operators\cite{bogdan}
\begin{equation}
b=\frac{1}{\sqrt 2}\left( \frac{d~}{dx}+\beta(x) \right) , \ \ \ 
b^*=\frac{1}{\sqrt 2}\left( -\frac{d~}{dx}+\beta(x) \right) ,  
\label{b*b}
\end{equation}
which, in order to satisfy the factorization $bb^*=H+\frac{1}{2}$, lead to a Riccati equation for the function $\beta(x)$
\begin{equation}
\beta' +\beta^2=1+x^2 .
\label{ricc-b}
\end{equation}
The immediate solution $\beta=x$ leads to the original anihilation/creation operators, while the general solution of the Riccati equation
\begin{equation}
\beta(x)=x+\phi(x)=x+\frac{e^{-x^2/2}}{\gamma+\int_0^x e^{-x'^2/2}dx'}
\label{betagen}
\end{equation}
leads to new hamiltonians (and potentials) defined by $b^*\,b\,=\tilde H-\frac{1}{2}$,
\begin{equation}
\tilde H=H-\phi'(x) \, ,  \ \ \ \tilde V(x)=\frac{x^2}{2}-\frac{d}{dx}\left[ 
\frac{e^{-x^2/2}}{\gamma+\int_0^x e^{-x'^2/2}dx'} \right] ,
\label{newpot}
\end{equation}
isospectral to the SHO potential, whose eigenfunctions are defined by
\begin{equation}
\tilde \psi_{n+1}=b^*\psi_n
\label{eigen2}
\end{equation}
for $n\geq 0$,  and with the ground state defined by the equation
\begin{equation}
b \tilde \psi_0=0 \, .
\label{tpsi0}
\end{equation}
The new potentials and eigenfunctions depend on the SUSY parameter $\gamma \in (\sqrt\pi/2,\infty)$.

In this article we would like to show that not all has been said about factorizing the hamiltonian (\ref{sho}), and that we can still find some hidden information through the factorization procedure.  We shall 
propose an alternative factorization, which includes both Mielnik's factorization and the original factorization, in terms of anihilation/creation operators, as particular cases, and which in its most general form would lead to a bi-parametric factorization of the hamiltonian.  
However, we shall only concentrate on a simple form of this factorization that leads to a new general equation for the SHO which includes its Schr\"odinger's equation and Hermite's equation as particular cases.

\section{ALTERNATIVE FACTORIZATION}

To begin with, let us introduce the couple of non-mutually adjoint operators
\begin{equation}
B=\frac{1}{\sqrt 2}\left( \frac{1}{\alpha(x)} \frac{d^2}{dx^2}+\beta(x) \right) , \ \ \ 
B^*=\frac{1}{\sqrt 2}\left( -\alpha(x) \frac{d^2}{dx^2}+\beta(x) \right) ,
\label{newb*b}
\end{equation}
and let us require again that they factorize the hamiltonian as $BB^*=H+\frac{1}{2}$.  Then, the functions $\alpha$ and $\beta$ have to satisfy the coupled equations
\begin{eqnarray}
\alpha' &+& \beta\alpha^2-\beta=0 \, ,
\label{coup-a}
\\
\beta' &+& \alpha\beta^2=(1+x^2)\, \alpha \, .
\label{coup-b}
\end{eqnarray}
These equations can be uncoupled by dividing the second one by $\alpha$, multiplying the first one by $\beta/\alpha^2$, and substracting, to obtain
\begin{equation}
\frac{d~}{dx} \left( \frac{\beta}{\alpha} \right) + \left( \frac{\beta}{\alpha} \right)^2 = 1+x^2 \, .
\label{ricc-ba}
\end{equation}
Clearly, Mielnik's solution (\ref{betagen}) is the general solution for this equation, introducing one parameter, $\gamma$.  A second parameter, $\delta$, will appear when we insert this solution into one of the equations (\ref{coup-a}) or (\ref{coup-b}) to find the complete biparametric solution.  When these parameters acquire the values that make $\alpha\equiv 1$ we shall recover Mielnik's factorization of the simple harmonic oscillator, just as when $\gamma\to\infty$ in Mielnik's factorization we recover the original factorization (\ref{a*a}).

Using Mielnik's solution (\ref{betagen}) it is easy to calculate the general solution to eqs.(\ref{coup-a},\ref{coup-b}), and to find the operators that factorize the SHO hamiltonian in terms of the product $BB^*$.  Then, we may consider carrying out the inverse product $B^*B$ to see where we are led to.
Obviously, we shall not obtain a newer Hamiltonian, due to the factors $\alpha^{-1}$ and $\alpha$ in (\ref{newb*b}), which may be the reason why this factorization was not paid attention to in the past.  What's more, it is obvious that this operator product will be lengthy and most easily won't give any new insight to the problem.  Therefore, we shall consider here only the most simple solution to eqs.(\ref{coup-a},\ref{coup-b}) to show that it leads to a new general equation for the SHO, lost in the trends of the SUSY factorization scheme.

\section{NEW GENERALIZED HERMITE EQUATION}

Let us now take the simplest solution to eq.(\ref{ricc-ba}), $\beta=\alpha\, x$.  Introducing this solution into eq.(\ref{coup-a}), we get a Bernoulli equation for $\alpha(x)$
\begin{equation}
\alpha' + x\alpha^3-x\alpha=0 \, .
\label{eqalpha}
\end{equation}
This equation is easily integrated, giving $\alpha$ and $\beta$ as
\begin{equation}
\alpha(x)=\frac{1}{\sqrt{1+\delta e^{-x^2}}} \ , \ \ \ \  
\beta(x)=\frac{x}{\sqrt{1+\delta e^{-x^2}}}  \, .
\label{alf-bet}
\end{equation}
To avoid singularities, we simply require that $0\leq \delta < \infty$.  

As we said before, the inverse operator product $B^*B$ will not give us a new Hamiltonian.  However, we can still introduce a second order operator defined by $\tilde {\cal L}=B^*B+1/2$,
\begin{equation}
\tilde{\cal L}= -\frac{1}{2}\frac{d^2~}{dx^2}+
\frac{\delta x e^{-x^2}}{1+\delta e^{-x^2}} \frac{d~}{dx}+
\frac{1}{2}\left[ \frac{x^2}{\left( 1+\delta e^{-x^2} \right)^2}-\frac{1}{1+\delta e^{-x^2}}+1 \right] .
\label{Ltilde}
\end{equation}
Also, defining the functions H$_n^\delta(x)$ by
\begin{equation}
\mbox{H}_{n+1}^\delta(x)=B^*\psi_n(x)
\label{unp1}
\end{equation}
we have that
\begin{equation}
\tilde{\cal L}\, \mbox{H}_{n+1}^\delta=\left(B^*B+\frac{1}{2}\right) \left( B^*\psi_n \right)=
B^*\left(BB^*+\frac{1}{2}\right) \psi_n=\left( E_n+1\right) \mbox{H}_{n+1}^\delta \, .
\label{unp1L}
\end{equation}

Now, as in SUSY QM, we can define the missing function H$_0^\delta(x)$ by requiring that 
$\tilde{\cal L}\mbox{H}_0^\delta=E_0\mbox{H}_0^\delta$, leading to the equation
\[
B\, \mbox{H}_0^\delta=\frac{1}{\sqrt 2}\left( \frac{1}{\alpha}\frac{d~}{dx}+\alpha x \right)\mbox{H}_0^\delta=0
\label{equ0}
\]
whose solution is
\begin{equation}
\mbox{H}_0^\delta=\alpha\psi_0 \, .
\label{u0}
\end{equation}
In fact, since $\beta=\alpha x$, eqs.(\ref{a*a}) and (\ref{newb*b}) tell us that $B^*=\alpha a^*$, and hence all (unnormalized) functions $\mbox{H}_{n}^\delta$ become $\mbox{H}_{n}^\delta(x)=\alpha(x)\,  \psi_n(x)$, $n\geq 0$.

Owing to the Sturm-Liouville theory,\cite{arfken} the functions $\mbox{H}_{n}^\delta$ are found to be orthogonal, since multiplying eq.(\ref{unp1L}) by the factor $-2\left( 1+\delta e^{-x^2} \right)$
we get the eigenvalue equation
\begin{equation}
{\cal L} \, \mbox{H}_{n}^\delta(x)+ E_n\, \omega(x)\, \mbox{H}_{n}^\delta(x)=0 \, ,
\label{unsl}
\end{equation}
where
\begin{equation}
{\cal L}= \left( 1+\delta e^{-x^2} \right) \frac{d^2~}{dx^2}-
2\delta x e^{-x^2} \frac{d~}{dx}-
\left[ \frac{x^2}{ 1+\delta e^{-x^2} }+ \delta e^{-x^2} \right]
\label{LL}
\end{equation}
is a new self-adjoint operator for the SHO, whose normalized eigenfunctions are 
\begin{equation}
\mbox{H}_{n}^\delta(x)=\left(\frac{1}{2^{n+1} n! \sqrt\pi}\right)^{\frac{1}{2}}
\left( e^{x^2}+\delta \right)^{-\frac{1}{2}} \, \mbox{H}_n(x) \, ,
\label{unfin}
\end{equation}
and where
\begin{equation}
\omega(x)=2\left( 1+\delta e^{-x^2} \right)
\label{peso}
\end{equation}
is the appropriate weight function.  Note that the eigenfunctions (\ref{unfin}) are not just the product of the Hermite polynomials and {\it any} other function, but a direct consequence of the factorization based on the operators $B,~ B^*$ (\ref{newb*b}).

\begin{figure}[ht]
\vspace*{2cm}
\begin{center}
\epsfig{figure=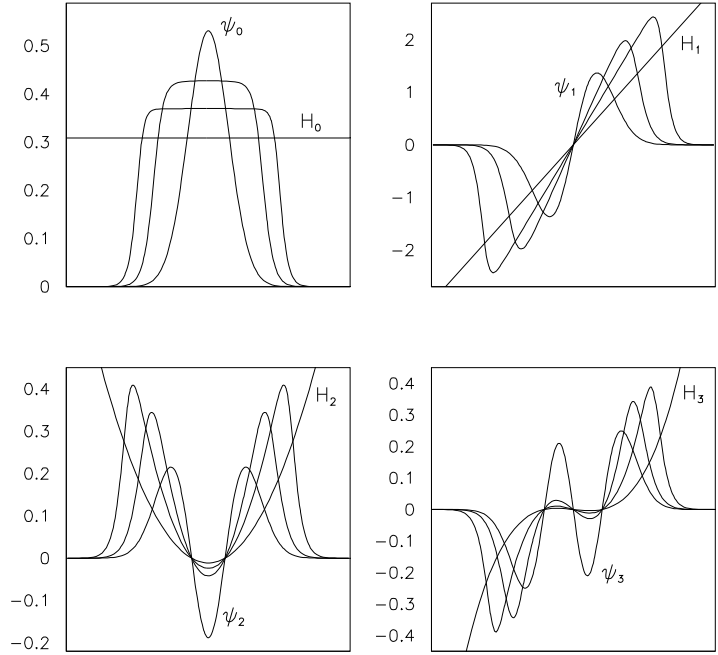, height=11cm}
\end{center}
\vspace*{-4cm}
\caption{The first four generalized Hermite functions 
H$_n^\delta(x)$ (arbitrary scale) for different values of  the parameter $\delta$.  When $\delta=0$ we get the quantum eigenfunctions $\psi_n(x)$, and for increasing  $\delta$, the eigenfunctions H$_n^\delta$ begin to acquire the form of the Hermite polynomials, H$_n(x)$, modulated  at $x\to \pm \infty$ by the vanishing exponential tails $e^{-x^2/2}$.} 
\end{figure}

The operator $\cal L$, the weight factor $\omega(x)$ and the eigenfunctions $\mbox{H}_{n}^\delta(x)$, all depend on the factorization parameter $\delta$.  In the case $\delta=0$, we have that $\alpha(x)\equiv 1$, and $B$ and 
$B^*$ become the original anihilation/creation operators (\ref{a*a}).  Also, the eigenvalue equation (\ref{unsl}) becomes Schr\"odinger's equation for the SHO, and $\mbox{H}_{n}^\delta(x)$ become the quantum eigenfunctions. 

On the other hand, in the limit $\delta\to \infty$, $\cal L$,  $\omega$, and $\mbox{H}_{n}^\delta$ become 
\begin{eqnarray*}
{\cal L}&\to& \delta e^{-x^2}  \frac{d^2~}{dx^2}-
2\delta x e^{-x^2} \frac{d~}{dx}- \delta e^{-x^2}  \, , \\
\omega &\to& 2 \delta e^{-x^2}  \, , \\
\mbox{H}_{n}^\delta &\to& \frac{1}{\sqrt\delta} \mbox{H}_n(x)  \, ,
\end{eqnarray*}
and the eigenvalue equation (\ref{unsl}) becomes Hermite's differential equation.  Therefore, we can call eq.(\ref{unsl}) a new {\it generalized Hermite equation} for the SHO, which includes Schr\"odinger's and Hermite's equations as particular cases for limiting values of the parameter $\delta$, linked by a continuos parametric transformation from one to the other.
This also allows us to call the eigenfunctions H$_n^\delta(x)$  new {\it generalized Hermite functions}.  

A plot of the first generalized Hermite  functions is shown in Fig.(1).  Notice the change in the functions behaviour as $\delta$ goes from zero to infinity, showing the continuos passage from the quantum mechanics eigenfunctions to the Hermite polynomials.

\section{RAISING AND LOWERING OPERATORS}

The corresponding raising and lowering operators, $c^*$, $c$, are easily deduced from the action of $a^*$ and $a$ over  $\psi_n(x)$.  The non-mutually adjoint operators
\begin{equation}
c^*=\alpha\, a^* \frac{1}{\alpha} = \frac{1}{\sqrt 2}\left( -\frac{d~}{dx}+
\frac{1+2\delta e^{-x^2}}{1+\delta e^{-x^2}} \, x\right)
\label{creaani}
\end{equation}
\begin{equation}
c=\alpha\, a \, \frac{1}{\alpha} = \frac{1}{\sqrt 2}\left( \frac{d~}{dx}+
\frac{x}{ 1+\delta e^{-x^2}} \right)
\label{creaani2}
\end{equation}
become the creation/anihilation operators of the quantum mechanics problem as $\delta\to 0$, and the raising and lowering operators for the Hermite polynomials as $\delta\to \infty$.  
They satisfy the relations
\begin{equation}
c^*\mbox{H}_n^\delta=\sqrt{n+1}\, \mbox{H}_{n+1}^\delta \ , \ \ \ \ c\, \mbox{H}_{n}^\delta=\sqrt{n}\, \mbox{H}_{n-1}^\delta \ ,
\label{cc*ops}
\end{equation}
which are the quantum eigenfunctions creation/anihilation operations when $\delta=1$, and become the raising/lowering operations for the Hermite polynomials in the limit $\delta\to\infty$.

They also satisfy that
\begin{eqnarray*}
cc^*\, \mbox{H}_{n}^\delta &=& \alpha\left(H+\frac{1}{2}\right)\frac{1}{\alpha} \, \mbox{H}_{n}^\delta=
(n+1)\, \mbox{H}_{n}^\delta \, , \\
c^*c\, \mbox{H}_{n}^\delta &=& \alpha\left(H-\frac{1}{2}\right)\frac{1}{\alpha} \, \mbox{H}_{n}^\delta=n\, 
\mbox{H}_{n}^\delta \, , 
\label{cc*act}
\end{eqnarray*}
and, therefore,
\begin{equation}
[c,c^*]=1 \, .
\label{cc*comm}
\end{equation}
Finally, due to the relation 
$a^*=(1/\alpha)B^*$ and eq.(\ref{unp1L}), 
$c^*$ and $c$ correspond to the creation/anihilation operators introduced by Mielnik\cite{bogdan}  as triple operator products.

\newpage
\section{ CONCLUSION}

In this article we have introduced an alternative factorization for the SHO hamiltonian that leads to the new self adjoint equation (\ref{unsl}), a parametric equation that becomes the SHO 
Schr\"odinger's equation in the limit $\delta\to 0$, and Hermite's equation in the limit $\delta \to\infty$.  
We therefore call this equation the new {\it generalized Hermite equation}
and the functions H$_n^\delta(x)$ (\ref{unfin}) the new  {\it generalized Hermite functions}.  
We have also shown that the raising and lowering operators (\ref{creaani}-\ref{creaani2}) correspond to the appropriate operators  in these limits. 
\\

\begin{acknowledgments}
We would like to acknowledge Erika Roldan for interesting and useful conversations. 
We also acknowledge support from CONACYT of Mexico, through a scholarship for MRG.
\end{acknowledgments}


\end{document}